\documentclass[twocolumn,aps,showpacs,showkeys]{revtex4}

\begin{document}

\title{Towards a new quantization of Dirac's monopole}

\author{A. I. Nesterov}
\email{nesterov@udgserv.cencar.udg.mx}
\homepage{http://udgserv.cencar.udg.mx/~nesterov}

\author{F. Aceves de la Cruz}
\email{fermin@udgphys.intranets.com}

\affiliation{Departamento de F{\'\i}sica, CUCEI, Universidad de
Guadalajara,\\
Guadalajara, Jalisco, M\'exico}

\begin{abstract}
There are several mathematical and physical reasons why Dirac's
quantization must hold. How far one can go without it remains an
open problem. The present work outlines a few steps in this
direction.
\end{abstract}


\maketitle

In his pioneer work \cite{Dir} Dirac showed that the existence of
just one magnetic monopole would explain electric charge
quantization. Nowadays this is known as {\it Dirac's quantization
rule}. In sixties Lipkin {\it et al} \cite{Lip} showed that in the
presence of the magnetic monopole the Jacobi identity for the
translational group failed. Later Jackiw \cite{Jac} found that the
Dirac's quantization restores the associativity of the finite
translations, but the infinitesimal generators remain
``nonassociative" and the Jacobi identity fails.

There exist various arguments based on the quantum mechanics,
theory of representations, topology and differential geometry in
behalf of the Dirac's rule \cite{Wu1,Wu2,Jac,GS,Nsh,K}. All of
them are related to the necessary conditions of saving the
situation and the problem of finding the sufficient conditions of
existing of magnetic point like monopole is still open. It is
therefore of interest to construct self consistent nonasociative
quantum mechanics with an arbitrary magnetic monopole charge or
maybe involving the distinct quantization rules related to the
magnetic monopole. The present work outlines a few steps in this
direction and the problems we have found.

Let us consider a point particle with electric charge $e$
and mass $m$ moving in the field of a magnetic monopole of charge $q$.
The non-relativistic classical Hamiltonian  may be written as follows
\cite{Gol}
\[
H = \frac{1}{2m r^2}({\mathbf p}\cdot{\mathbf r})^2 +
\frac{1}{2mr^2}({\mathbf J}^2 - \mu^2)
\]
where
\begin{equation}
{\mathbf J} = {\mathbf r}\times({\mathbf p} - e{\mathbf A}) -
\mu\frac{\mathbf r}{r}, \quad \mu = eq
\label{eq1}
\end{equation}
is the conserved total angular momentum and we set $\hbar = c =1$.
The last term in the formula (\ref{eq1}) usually is interpreted as the
contribution of the electromagnetic field \cite{Gol,Wil,Gol2,Lyn}, which
carries an angular momentum
\[
{\mathbf L}_{em}=\frac{1}{4\pi}\int {\mathbf r} \times({\mathbf
E}\times{\mathbf B}) d^3 r = - \mu\frac{\mathbf r}{r}
\]
Also this is known as {\it Poincar\'e magnetic angular momentum}
\cite{Ber}. A magnetic field of the Dirac's monopole is
\begin{equation}
{\mathbf B} = q \frac{\mathbf r}{r^3}
\label{eq_0}
\end{equation}
and any choice of the vector potential $\mathbf A$ being compatible with
(\ref{eq_0}) must have singularities, the so-called {\it Dirac's string}.

At the quantum level the operator
\begin{equation}
{\mathbf J} = {\mathbf r} \times \left({-i\mathbf \nabla} - e{\mathbf
A}\right) - \mu\frac{\mathbf r}{r}
\end{equation}
representing the angular momentum $\mathbf J$ has the same
properties as a standard angular momentum and obeys the following
commutation relations
\begin{eqnarray}
&&[H, {\mathbf J}^2] = 0, \quad [H, J_i] = 0,\quad  [{\mathbf J}^2, J_i] =
0 \label{eq5a} \\
&&[J_i, J_j] = i\epsilon_{ijk}J_k,
\end{eqnarray}

Choosing the vector potential as
\[
{\mathbf A} = -q\frac{1 + \cos\theta}{r\sin\theta}\;{\hat{\mathbf
e}}_{\varphi}.
\]
we find
\begin{eqnarray}
&&J_{\pm}= e^{\pm i\varphi}\bigg(\pm\frac{\partial}{\partial\theta}
+i\cot\theta \frac{\partial}{\partial\varphi} -
\mu\frac{1+\cos\theta}{\sin\theta} \bigg),\\
&&J_0=-i\frac{\partial}{\partial\varphi} + \mu,\\
&&{\mathbf J^2} =-\frac{1}{\sin{\theta}}
\frac{\partial~}{\partial\theta}\left(\sin{\theta}
\frac{\partial~}{\partial\theta}\right) -
\frac{1}{\sin^2{\theta}}\frac{\partial^2~}{\partial\varphi^2} - \nonumber\\
&&-i\frac{2\mu}{1 -\cos{\theta}}\frac{\partial~}{\partial\varphi}
+\mu^2\frac{1 + \cos{\theta}}{1 - \cos{\theta}} +\mu^2
\end{eqnarray}
where  $J_{\pm} = J_x \pm iJ_y$ are the raising and the lowering
operators for $J_0 = J_z$ satisfying the standard commutations
relations
\[
[J_0,J_{\pm}] = \pm J_{\pm}, \quad [J_{+},J_{-}] = 2J_{0}.
\]

It is known that for the Dirac's monopole problem the Schr\"odinger
equation written in the spherical coordinates admits the separation of
variables and the eigenfunctions of the angular part of the Hamiltonian are
\cite{Wu2}
\begin{equation}
Y_{jn\mu}(\theta, \varphi) = \frac{1}{\sqrt{2\pi}}\mbox{e}^{i(j -
\mu - n)\varphi}P_{jn\mu}(\cos\theta),
\label{eq2}
\end{equation}
with $n$ as a non-negative integer and
\begin{eqnarray}
&&P_{jn\mu}(u) = \frac{1}{\sqrt{h_{jn\mu}}}(1 -
u)^{\alpha/2}(1 + u)^{\beta/2}P_{n}^{(\alpha, \beta)}(u),\nonumber \\
&&h_{jn\mu} = \frac{2^{2(j - n) + 1}}{2j + 1}\frac{\Gamma(j + \mu
+ 1)\Gamma(j - \mu + 1)}{n!\Gamma(2j - n + 1)}, \nonumber \\
&&\alpha = j + \mu - n, \quad\beta = j - \mu - n, \nonumber
\label{eq3}
\end{eqnarray}
here $\Gamma$ is the Gamma function, $j(j + 1)$ and $j - n$ are the
eigenvalues of ${\mathbf J}^2$ and $J_z$ respectively, and $P_{n}^{(\alpha,
\beta)}$ are the Jacobi polynomials.

Using the rotational invariance of the system and restricting by
the unitary representations of the group SO(3) one can obtain the
Dirac's quantization rule $2\mu=$ integer. This results in the
following: {\it the only way to avoid Dirac's rule is to consider
a nonunitary representation of the rotation group}.

Further we will try to describe the charge-monopole system as a free
particle with an arbitrary spin, the so-called {\it anyons}, and relate
this with the nonunitary representations of the rotation group. The idea
to describe Dirac's monopole as a particle with a spin is not new
\cite{Gol,Gol2,Ply}. However, in the known models anyon's statistics is
based on the SO(2,1)-symmetry \cite{Ply,Jan}. and it is not clear how
anyons may be appeared in a non-relativistic theory and its relation with
the group SO(3).

Following \cite{Ply} we consider the Lagrangian describing a
non-relativistic charged particle in the field of a magnetic monopole
\[
L= \frac{m}{2}\dot{\mathbf r}^2 + e{\mathbf A}\cdot\dot{\mathbf r}
\]
where $\mathbf A$ is the vector potential determining the magnetic field
(\ref{eq_0}). In the limit $m \rightarrow 0$, the action is given by the
integral of the one-form $\theta = eA_i dx^i$ and we have
\begin{equation}
d\theta = -\frac{1}{2{\mathbf
S}^2}\varepsilon_{ijk}S_idS_j\wedge{dS_k}, \quad {\mathbf S}
= -\mu\frac{\mathbf r}{r}
\label{eq4}
\end{equation}
The two-form $d\theta$ is closed and nondegenerate, and is interpreted as
a symplectic form corresponding to the symplectic potential $\theta$
defined on the unit two-sphere. It can be used to introduce
Poisson brackets on $S^2$ as follows
\begin{equation}
\{S_i, S_j\} = \varepsilon_{ijk}S_k,
\label{eq5}
\end{equation}
and thus the angular momentum of the electromagnetic field
$\mathbf L_{em}$ can be considered as a {\it classical spin}.

Developing this idea, one can give the alternative description of
the charge-monopole system as a free particle of the fixed spin
with translational and spin degrees of freedom. The corresponding
constrained Hamiltonian is given by
\begin{eqnarray}
H= \frac{1}{2m}\bigl(\mathbf p  - \frac{1}{r^2}(\mathbf r \times
{\mathbf S})\bigr)^2 + \lambda(\mathbf S \cdot \mathbf r + \mu r)
\label{ham}
\end{eqnarray}
where the second term describes the spin-orbit interaction. On the
constraint surface ${\mathbf S}\cdot{\mathbf r} + \mu{r} \approx 0$
the dynamics generated by the Hamiltonian (\ref{ham}) is exactly the same
as the dynamics of the initial charge-monopole system \cite{Ply}.

The above considerations is based on the conventional approach to
the spin and therefore involves implicitly the Dirac's
quantization rule. The case of $\mu$ being not necessarily integer
or half integer implies making use of the nonunitary
representations of the group SO(3).

Let the set of $\{S_{\pm},S_0\}$ and  $\{L_{\pm},L_0\}$ form the
algebra $so(2,1 )$ and $so(3)$ respectively
\begin{eqnarray}
{[}S_{+}, S_{-}] = -2S_0, \quad {[}S_{\pm}, S_0] = {\mp}S_{\pm} \\
{[}L_{+}, L_{-}] = 2L_0, \quad {[}L_{\pm}, L_0] = {\mp}L_{\pm}.
\end{eqnarray}
We introduce a direct sum $so(3)\oplus so(2,1)$ as follows
\begin{equation}
J_0 = L_0 + S_0, \; J_{+} = L_{+} - S_{+}, \; J_{-} = L_{-} + S_{-}.
\label{eq6}\\
\end{equation}
The computation shows that the operator $\mathbf J$ obeys the
commutation relations of the algebra $so(3)$
\[
[J_0,J_{\pm}] = \pm J_{\pm}, \quad [J_{+},J_{-}] = 2J_{0}
\]
and identifying $J_i$'s as
\begin{eqnarray}
J_{1} = L_{1} + iS_{2}, \; J_{2}
= L_{2} + iS_{1},\; J_3=J_0
\end{eqnarray}
we find
\[
[J_i,J_j] = i\varepsilon_{ijk} J_k.
\]
Notice that the group $\rm SO(3)\otimes SO(2,1)$ has previously
appeared in Dirac monopole theory, as a dynamical invariance group
\cite{Jac1}.

We construct the representation by using the generators of the
group $SO(3)$ with the standard action on the state $|l,\tilde m
\rangle$ given by
\begin{eqnarray*}
&&{\mathbf L}^2|l, \tilde{m}\rangle = l(l + 1)|l, \tilde{m}\rangle, \\
&&L_0|l, \tilde{m}\rangle = \tilde{m}|l, \tilde{m}\rangle,\\
&&L_{+}|l,
\tilde{m}\rangle = \sqrt{(l - \tilde{m})(l + \tilde{m} + 1)}|l, \tilde{m} +
1\rangle,\\
&&L_{-}|l, \tilde{m}\rangle = \sqrt{(l + \tilde{m})(l -
\tilde{m} + 1)}|l, \tilde{m} - 1\rangle,
\end{eqnarray*}
and the unitary infinite dimensional representation of the group
$SU(1, 1)$ relating to the generators $S_i$'s.

For the representation bounded above we have \cite{Jan}
\begin{eqnarray*}
&&{\mathbf S}^2|\lambda, n\rangle = \lambda(\lambda - 1)|\lambda, n\rangle,\\
&&S_0|\lambda, n\rangle = -(\lambda + n)|\lambda, n\rangle,\\
&&S_{+}|\lambda, n\rangle = -\sqrt{(2\lambda + n - 1)n}|\lambda, n
-1\rangle,\\
&&S_{-}|\lambda, n\rangle = -\sqrt{(2\lambda + n)(n + 1)}|\lambda,
n + 1\rangle,
\end{eqnarray*}
where $n=0,1,2,\dots,\infty$ and $\lambda$ is an arbitrary
parameter, and for the highest-weight state $|\lambda,0\rangle$
one obtains
\begin{eqnarray*}
S_0|\lambda, 0\rangle = -\lambda |\lambda, 0\rangle,\quad
S_{+}|\lambda, 0\rangle = 0.
\end{eqnarray*}
Thus $\lambda$ is the eigenvalue of $J_0$ for the state
$|\lambda,0\rangle$ and characterizes the representation. Recently
the infinite dimensional representations of the group SO(2,1) have
been used for the description of a theory of anyons where the spin
of the one-particle states is taken to be $s= 1-\lambda$
\cite{Jan} (see also \cite{Ply1,Ply2}).

Starting with the highest-weight state $|j,j\rangle = |l,l\rangle\otimes
|\lambda,0\rangle$ corresponding to the eigenvalue $l-\lambda$ of the
operator $J_0$ and applying the lowering operator $J_{-}$  one can
construct all states as follows
\begin{eqnarray}
&&J_{-}|j, m\rangle = \sqrt{(j + m)(j - m + 1)}|j, m - 1\rangle,\\
&&J_{+}|j, m\rangle = \sqrt{(j - m)(j + m + 1)}|j, m - 1\rangle, \\
&&m= j, j-1, \dots, -\infty \nonumber
\label{rep}
\end{eqnarray}

It is easy to see that the arised representation is nonunitary
infinite dimensional and bounded above by $m \le j$, instead of
$-j \le m \le j$ well known  for the unitary finite-dimensional
representations of the rotation group.

The analysis of the nonunitary representation shows that the
allowed values of $j,\mu$ are $ \mu^2 \le j(j + 1)$ and  $j - \mu$
= integer, which is an alternative choice to the Dirac rule
\cite{Nes}.

In our model the charge-monopole system is interpreted as a free
anyon with translational and spin degrees of freedom. The close
approach has been applied in \cite{Ply3,Ply4} for description of a
fractional spin in $(2+1)-$ dimensions.

Recently Mart{\'\i}nez-y-Romero {\it et al} \cite{Mar, Mar1} have
used a nonunitary representation for a Dirac particle in a
Coulomb-like field and Davis and Ghandour \cite{Dav} have showed
that nonunitary transformations are not so strange in quantum
mechanics. The physical consequences of using nonunitary
representations in Dirac's monopole problem are not clear till now
and the work is in progress.

\acknowledgments

We would like to thank R. Jackiw and M. Plyushchay for helpful
comments on Dirac monopole theory and related subjects.

\end{document}